\documentclass[5p,number,times]{elsarticle}

\usepackage{graphicx}
\usepackage{latexsym}

\newcommand{\greeksym}[1]{{\usefont{U}{psy}{m}{n}#1}}
\newcommand{\umu}{\mbox{\greeksym{m}}}

\newcommand{\uOmega}{\mbox{\greeksym{W}}}

\newcommand{\mrm}{\mathrm}
\newcommand{\Neq}{\mrm{n}_{\mrm{eq}}/\mrm{cm}^{2}}

\begin{document}
\begin{frontmatter}
\title{Radiation hardness of CMS pixel barrel modules}
\author[psi]{T.~Rohe\corref{cor}}
\ead{tilman.rohe@cern.ch}

\author[ku]{A.~Bean}
\author[psi]{W.~Erdmann}
\author[psi]{H.-C.~K\"astli}
\author[uic]{S.~Khalatyan}
\author[psi]{B.~Meier}
\author[ku]{V.~Radicci}
\author[psi,ku]{J.~Sibille}

\address[psi]{Paul Scherrer Institut, Villigen and W\"urenlingen, Switzerland}
\address[ku]{University of Kansas, Laurence KS, USA}
\address[uic]{University of Illinois at Chicago, USA}

\cortext[cor]{Correponding author}

\begin{abstract}
Pixel detectors are used in the innermost part of the multi purpose experiments at LHC 
and are therefore exposed to the highest fluences of ionising radiation, 
which in this part of the detectors consists mainly of charged pions. The radiation hardness of all 
detector components has thoroughly been tested up to the fluences expected at the LHC. In case of an LHC upgrade, the fluence will be much higher 
and it is not yet clear how long the present pixel modules will stay operative in such a harsh environment.
The aim of this study was to establish such a limit as a benchmark
for other possible detector concepts considered for the upgrade.

As the sensors and the readout chip are the parts most sensitive to radiation damage,
samples consisting of a small pixel sensor bump-bonded to a CMS-readout
chip (PSI46V2.1) have been irradiated with positive 200\,MeV pions at PSI up to 
$6\times 10^{14}\,\Neq$ and with 21\,GeV protons at CERN up to $5\times 10^{15}\,\Neq$. 

After irradiation the response of the system to beta particles from a $^{90}$Sr source was 
measured to characterise the charge collection efficiency of the sensor. Radiation induced
changes in the readout chip were also measured. The results show that the present 
pixel modules can be expected to be still operational after a fluence of $2.8\times 10^{15}\,\Neq$.
Samples irradiated up to $5\times 10^{15}\,\Neq$ still see the beta particles. However,
further tests are needed to confirm whether a stable operation with high
particle detection efficiency is possible after such a high fluence.
\end{abstract}

\end{frontmatter}

\section{Introduction}

Hybrid silicon pixel detectors are used in the innermost part of the multi purpose experiments at 
LHC at distances between 4~and 12\,cm from the beam \cite{ref:cms,ref:atlas}. They are therefore 
exposed to the highest fluences of ionising radiation, which in this part of the detectors consists 
mainly of charged pions. The radiation hardness of all detector components has thoroughly been tested 
up to the fluences expected at the LHC \cite{ref:tb-03-04}. A replacement of the present CMS~pixel
detector  by an improved 4-layer system is already planned for 2014 \cite{ref:alice}. 
As a two stage luminosity upgrade of the LHC is foreseen shortly after \cite{ref:slhc}, it has 
to be assured that  the new system is suitable for peak luminosities above 
 $10^{34}\,$cm$^{-2}$s$^{-1}$ in terms of radiation hardness and data transfer.
The short time scale allows only for very limited  R\&D and it has to be 
tested if the present detector concept can be adjusted to the new specifications. 
This paper reports on the radiation hardness of both the sensor and the readout chip, with 
the aim to estimate their fluence limits and their lifetime in an upgraded LHC.
 
\section{CMS Modules}

The pixel detector of CMS is a so called {\em hybrid} pixel detector. This means
that the sensor  part, which actually detects the charged particles, and the 
readout electronics are on separate substrates. Each sensing element (pixel) is
connected to its preamplifier located on the readout chip by a small indium bowl, 
the so-called bump. The power and I/O pads of the readout chip are then wire bonded 
to a high density interconnect (HDI) which provides the connection to the outside.

The two parts most sensitive to radiation damage are the sensor and the readout chip. 
Therefore this study is limited to those components. In order to further 
simplify the irradiation procedure only small modules consisting of one ROC and a small 
sensor covering the area  of this ROC were used. Such samples are about
$1\times 1$\,cm$^{2}$ large which is about the beam spot size in the irradiation 
sites chosen.  So, scanning during irradiation could  be avoided. However, in
contrast to strip detectors, where the strip length is important, small 
pixel devices contain all relevant features of a full module and the results
obtained in this study are fully valid.

\subsection{The Sensor}

The sensors for the CMS pixel barrel are made of silicon and 
follow the so called ``n-in-n'' approach. 
The collection of electrons is advantageous because of their 
higher mobility compared to holes. It causes a larger Lorentz
drift of the signal charges which leads to charge sharing between 
between neighboring pixels and so improves the spatial resolution.
Further the higher mobility of electrons makes them less
prone to trapping, which is essential after high fluences of
charged particles.  After irradiation induced space charge sign inversion, 
the highest  electric field in the sensor is located close to the n-electrodes 
used to collect the charge, which is also of advantage.

The choice of n-substrate requires a double sided sensor process,
meaning that both sides of the sensor need photolithographic steps. This
leads to higher costs compared to single sided p-in-n (or n-in-p) sensors.
However, the double sided sensors have a guard ring scheme
where all sensor edges are at a ground potential and this
greatly simplified the design of detector modules.  The n-side isolation was
implemented through the so-called moderated p-spray technique~\cite{ref:mod}
with a punch through biasing grid.

The sensor samples were taken from wafers of the main production run for the
CMS pixel barrel which were processed on approximately $285\,\umu$m thick n-doped 
DOFZ silicon according to the
recommendation of the ROSE-collaboration~\cite{ref:rose}. The resistance of the
material prior to irradiation was $3.7\,\mrm{k}\uOmega\mrm{cm}$ leading
to an initial full depletion voltage of $V_{\mrm{fd}} \approx 55\,$V.


\subsection{The Readout Chip}

The CMS pixel readout chip (called PSI46V2.1) described in detail
in \cite{ref:roc} has been fabricated in 
a commercial five-metal layer $0.25\,\umu$m process available via
CERN. The chip contains about 1.3 million transistors on an area 
of $7.9 \times 9.8\,$mm$^{2}$. It is divided into three functional 
blocks: the array of $80\times 52$ pixel unit cells of 
$100\times 150\,\umu$m$^{2}$ size organised in 
26~double columns, the double column perifery, and a control and 
a chip perifery containing a supply and control block.

The radiation hardness of the $0.25\,\umu$m technology is  
well known if certain layout recommendations are followed \cite{ref:radhardroc}. 
This concerns the leakage currents, which are strongly surpressed by
a ``closed gate'' geometry of the NMOS transistors and channel stop rings. 
The shift of the transistor thresholds is small due to the small thickness of the
gate oxides in such processes. Other radiation induced effects
like the reduction of the surface mobility and the minority
carrier life time remain. The former might effect the transconductance
of the transistors and by this the speed of the chip while the latter 
influences the bipolar devices like diodes.

\section{Sample Preparation}

The readout chips taken are identical to those used for building the CMS
pixel detector. The small sensors were produced on the same wafers as
those for full modules and identical to those apart from the size.
Both parts were joined using the bump bonding process used for the 
module production for the CMS pixel barrel \cite{ref:bump}.
As this procedure includes processing steps at temperatures
above $200^{\circ}\,$C, it was done before irradiation.
This means that the sensors  and  readout chips had to be irradiated at the same 
time.   This way, a realistic picture of the situation in CMS 
after a few years of running could be obtained.

The sandwiches of sensor and readout chip were irradiated at the PSI-PiE1-beam 
line \cite{ref:piE1}
with positive pions of momentum 200\,MeV/c to fluences up to
$6\times 10^{14}\,\Neq$ and with 26\,GeV/c protons at CERN-PS \cite{ref:irrad1} up to
$5\times 10^{15}\,\Neq$.

All irradiated samples were kept in a commercial freezer at $-18^{\circ}\,$C after
irradiation. However, the pion irradiated ones were accidentally warmed up to room
temperature for a period of a few weeks.
 
\section{Sensor Performance}
The aim of this study was to determine the signal obtained from minimum
ionizing particles (m.i.p.) which is the limiting factor for the use of such
type of sensors.

\subsection{Measurement Procedure}

For reasons of simplicity, a $^{90}$Sr source has been used for inducing the signal
in the sensor. The $\beta$-spectrum of the daughter decay of $^{90}$Y has 
an endpoint energy of about 2.3\,MeV and therefore contains particles
which approximate a m.i.p. well. However, there are also a lot of low energy particles 
which lead to an energy deposition in the sensor which is much higher. 

The samples were mounted on a pair of water cooled Peltier elements and
kept below $-10^{\circ}\,$C. The source was placed inside the box a 
few millimetres above the sample. For acquiring the data a so-called
{\em random trigger} was used. In this mode of operation an arbitrary
cycle of the  clock is stretched by a large factor, and a trigger is sent in a
way that all the hits within this clock cycle are read out.
The stretch factor was a adjusted such that about 80\,\% of the triggers
showed hits. As this method does not allow for hit detection efficiency
measurements, the upgrade of the setup with an independent scintillator
trigger is currently underway.

The testing and calibration procedure was similar to what is used for
the qualification of CMS pixel barrel modules \cite{ref:andrey}. 
The threshold to which the pixels were adjusted was $4000\,$e$^{-}$ 
for all samples up to a fluence of $1.2\times 10^{15}\,\Neq$. For the higher 
fluence the threshold was lowered to a value between $2000$ and $3000\,$e$^{-}$.
After, calibration data was taken using the $^{90}$Sr source. The sensor
bias was varied from 50\,V up to 1100\,V for some of the most irradiated samples.
The temperature was kept stable within $0.2^{\circ}\,$C during the bias
scan.

The data was analysed offline. First, all analog signals were converted
into an absolute charge value \cite{ref:sarah}. Then a mask was generated
to exclude noisy and unbonded pixels from further analysis. Finally, all
clusters not contiguous with a masked pixel were reconstructed. The
charge of the clusters was histogrammed separately for each cluster size.
The spectra were fit to a Landau function convoluted with a Gaussian.
The charge value quoted is the most probable value (MPV) from this fit.

\begin{figure}[t]
\centering
\includegraphics[width=.98\linewidth]{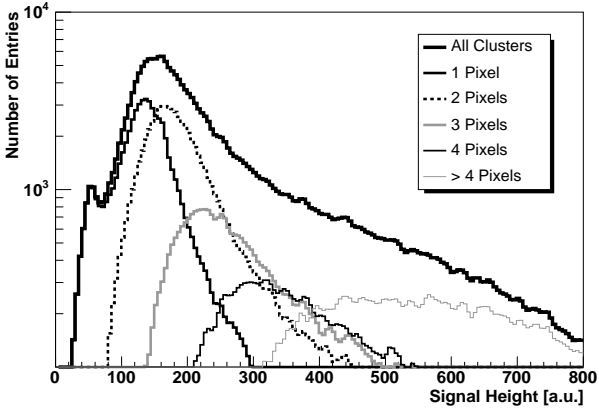}
\caption{Pulse height distribution of a sample irradiated to
 $2.8\times 10^{15}\,\Neq$ at a bias voltage of 800\,V in 
 arbitrary units (1 unit corresponds to about 65 electrons)\label{fig:spectum3E15}}
\end{figure}

The radiation of the $^{90}$Sr source contains a sizable fraction of
low energy electrons which might be stopped in the sensor and cause a 
much higher signal than a minimum ionising particle. A part of the 
secondary electrons, however, travels in parallel to the sensor surface
resulting in large clusters. Figure~\ref{fig:spectum3E15} shows the
pulse height spectrum of a sample irradiated to $2.8\times 10^{15}\,\Neq$ 
at a bias voltage of 800\,V for different cluster sizes. The large fraction
of 2-clusters and the occurrence of clusters larger than 2 pixels cannot
be explained by simple charge sharing and clearly indicated the presence
of low energy electrons from the source. The same is the case for the strong 
dependence of the MPV on the cluster size. In order not to bias
the data only the spectra of single hit clusters have been used.

\subsection{Results}

\begin{figure}[t]
\centering
\includegraphics[width=.98\linewidth]{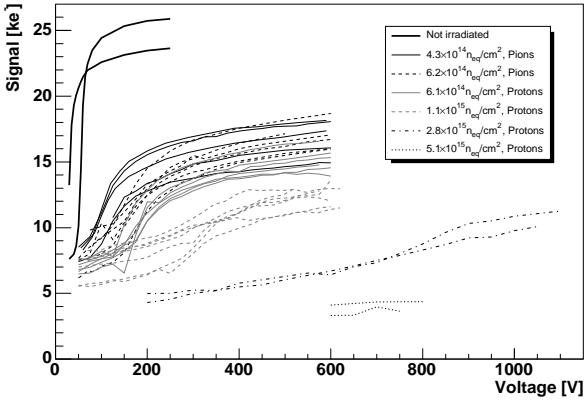}
\caption{Most probable signal from single pixel clusters as a function of the sensor
  bias. Each line represents one sample \label{fig:q-v}}
\end{figure}

\begin{figure}[t]
\centering
\includegraphics[width=.98\linewidth]{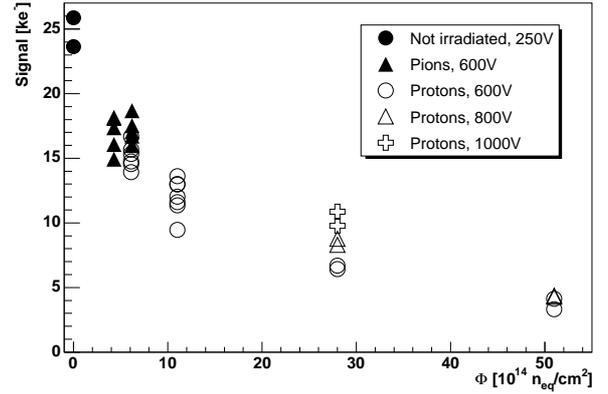}
\caption{Most probable signal as a function of the irradiation fluence \label{fig:q-phi}}
\end{figure}

Figure~\ref{fig:q-v} shows the values of the MPV for all the measured samples and
all biases. For the unirradiated samples the steep rise of the signal at the full 
depletion voltage
of $V_{\mrm{depl}}\approx 55\,$V is visible. The signal plateau is reached very fast.

The samples irradiated to fluences in the $10^{14}\,\Neq$-range also show a saturation
of the signal above roughly 300\,V. However, the variation of the saturated signal
for the same 
irradiation fluence is of the same order of magnitude as the reduction of the
signal from $4.3$ to $6.2\times 10^{14}\,\Neq$. A distinction of
the irradiation fluence of the samples from the height of the plateau is not possible.
This strong variation cannot be explained by differences in the sensor thickness and
is probably caused by the imperfection of the pulse height calibration. It relies 
on the assumption that the injection mechanism 
for test pulses is equal for all readout chips, which is not the case. Variations of the
injection capacitor are larger than $15\,\%$.

The differences in the irradiation fluence are better reflected
in the ``low'' voltage
part of the curves, where the voltage from which the signal starts to rise
reflects the radiation-induced increase of the space charge.

For the samples irradiated to fluences above $10^{15}\,\Neq$, no saturation of the
signal with increasing bias is visible. It is remarkable that even after 
a fluence of $2.8\times 10^{15}\,\Neq$ a charge of more than 10\,000~electrons can
be collected if a bias voltage over 800\,V is applied. 

The samples irradiated to $5.1\times 10^{15}\,\Neq$, can clearly detect
the particles. However, the signal obtained is well below 5000~electrons
which is too little for reliable operation as a tracking device. 
The data were taken
at a temperature of about $-15^{\circ}\,$C and therefore the
leakage current was above $50\,\umu$A for the about $1\,$cm$^{2}$ large
device. This did not allow measurements above 800\,V. 

To display the development of the signal height as a function of
the fluence, the charge at 600\,V was measured for each sample (250\,V for 
the unirradiated ones) and plotted in Fig.~\ref{fig:q-phi}. In addition, 
values for 800\,V and 1000\,V are shown for the highest fluence.
Apart from the large fluctuations
the reduction of the charge with fluence is
visible. Further it becomes obvious that it pays to go to 
very high bias voltages  if the fluences exceed  $10^{15}\,\Neq$.

The charge values in this study are lower than those reported
elsewhere \cite{ref:liv,ref:lj}. 
The authors in these studies suggested that some kind of avalanche
multiplication was
the reason for the higher than expected signals.
The sensors used in this study are designed differently. The layout 
of the pixel's n$^{+}$-implants was optimized to avoid peaks of the electric
field. Therefore the IV-curves of the sensors tested do not show
evidence for an onset of micro discharge. 
 
\section{Performance of the Readout Chip}

In general it can be stated that all of the ROCs, even at the
highest fluence, worked well and could be operated using the 
standard calibration software. For the ROC with a fluence
above $2\times 10^{15}\,\Neq$ the feedback circuit of the 
preamplifier and shaper had to be adjusted, and the sampling point 
of the ADC, reading the chip data had to be shifted for a few nanoseconds.

The parameters of the readout chip which are subject to change due
to irradiation are not directly accessible in the readout chips.
For an extraction of the transistor parameters a special test chip would be
necessary. However, there are some chip parameters which 
give, indirectly, hints on the radiation induced changes in the ROC.

\begin{figure}[t]
\centering
\includegraphics[width=.98\linewidth]{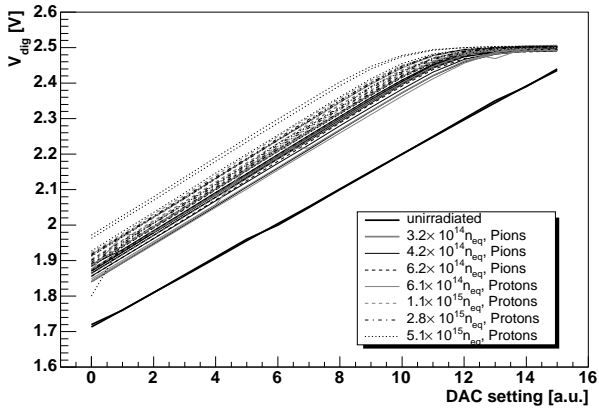}
\caption{Output voltage of the ROC-DAC ,,$V_{\mrm{dig}}$'' for two unirradiated
   chips and irradiated samples up to a fluence of $5.1\times 10^{15}\,\Neq$ \label{fig:Vdig}}
\end{figure}

One very sensitive block in the periphery is the band gap reference which is used
as a reference for all internal voltages set. One of theses voltages is
stabilized using an external SMD-capacitor and can therefore be measured 
using a probe. The band gap reference is in principle a forward biased diode
whose reference voltage might be dependent on the minority carrier life time.
Measuring the digital voltage~$V_{\mrm{dig}}$ as a function of the adjusted
DAC setting is therefore a test of the stability of the band gap reference.
The result of this measurement is shown in Fig.~\ref{fig:Vdig}. The 
output of the band gap reference shows a shift of about 10\,\% already
at the lowest fluence and then stays constant. This is a hint of a surface
effect which usually saturates at low doses. The saturation of the voltage
at DAC values above 10 is due to the fact the supply voltage of 2.5\,V
cannot be exceeded.

\begin{figure}[t]
\centering
\includegraphics[width=.98\linewidth]{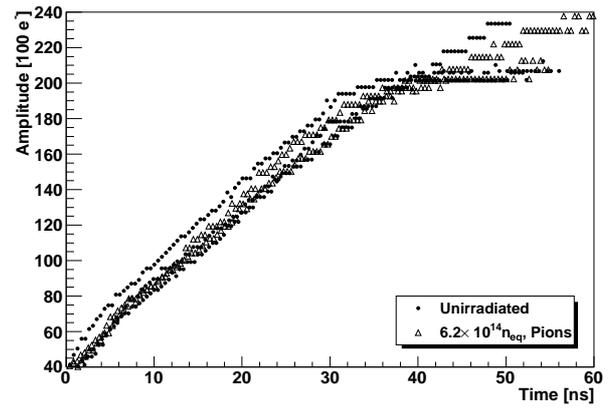}
\caption{Pulse shape of an injected test signal measured in one pixel of
   three unirradiated ROCs and two ROCs irradiated to $6.2\times 10^{14}\,\Neq$ \label{fig:pulse}}
\end{figure}

Another parameter, which might be prone to radiation induced changes, is the
rise time of the preamplifier. This causes the so called {\em time walk}
and either needs to be compensated by an increase of the analog
current or results in a higher in-time threshold. 

The pulse shape 
of the PSI46V2.1 cannot be measured directly, but an indirect method gives
a hint on its leading edge. It is based on calibration pulses which
can be injected into each pixel. 
With a certain DAC setting, called {\em calibration delay}, the time 
of this pulse can be shifted with respect to the leading edge of the 
clock cycle whose signals will be read out. If this DAC is scanned while
the height of the calibration signal and the threshold stays constant,
an efficiency curve will be obtained. If the pulse is injected too early
it will cross the threshold in the clock cycle before the readout
and will not be seen. If it is too late it will be assigned to the next
clock cycle. The plateau of full efficiency in-between the two regions
is exactly one clock cycle wide i.e. 25\,ns. If this scan is repeated with
a lower threshold and the same calibration pulse height, the efficiency plateau
will be slightly shifted due to the fact that the threshold will be 
reached slightly earlier. If the whole space of the threshold and the 
calibration delay is scanned, the timing of the leading edge of the
pulse shape can be reconstructed. 

The result of such a reconstruction is shown in fig.~\ref{fig:pulse}
for one pixel each in three unirradiated ROCs and two ROCs irradiated 
to $6.2\times 10^{14}\,\Neq$. The analog current was the same in all
the measurements. The rise time of the pulse is about 35\,ns and,
most remarkable, it does not change with irradiation, at least not 
up to the level investigated.

\section{Conclusion}

Sandwiches of a CMS pixel readout chip and a small single chip
sensor have been irradiated up to flucences of $5.1\times 10^{15}\,\Neq$.
Test with a $^{90}$Sr source showed a signal above 10\,000 electrons 
per m.i.p. can be achived at bias voltage of or below 600\,V up to a
fluence of about  $10^{15}\,\Neq$. The samples irradiated up to 
$2.8\times 10^{15}\,\Neq$, however, needed a bias voltage of 1000\,V to reach
a signal of 10\,000 electrons. This indicates the suitabiliy of these 
devices for an upgraded LHC up to rather small radii.
The samples which obtained a fluence
of $5.1\times 10^{15}\,\Neq$ could only be measured up to a bias voltage 
of 800\,V due to limitations of the cooling. They still showed signals, but
their amplitude seems too low for an eficient tracking. Measurements 
at higher bias voltages are a subject of further studies, as well as
efficiency measurements using an independent scintillator trigger.

The readout chip performed well even after the highest fluence. It could be
operated with the standard calibration method and only very few settings,
like the feedback resistor of the preamplifier and shaper, needed to be
adjusted. A degradation of the amplifier's rise time could not
be detected by this study. The band gap reference of the chip drifted by
about 10\,\% already after lowest fluence, but this shift saturated.
Further studies to characterize the irradiated readout chips are under way.

\section*{Acknowledgement}

The pion irradiation at PSI would not have been possible without
the beam line support by D.~Renker and K.~Deiters, PSI, the
logistics provided by M.~Glaser, CERN, and the great effort
of C.~Betancourt and M.~Gerling, UC Santa Cruz (both were supported by
a financial contribution of RD50 and PSI).

The proton irradiation was carried out at the CERN irradiation facility.
The authors would like to thank M.~Glaser and the CERN team for the outstanding
service.

Thanks are also due to N.~Krzyzanowski and E.~Stachura, UIC for the measurements of
the irradiated ROCs.

The work of A.~Bean and V.~Radicci is supported by the PIRE grant OISE-0730173 of the US-NSF.

The sensors  were produced by CiS GmbH in Erfurt, Germany.

\bibliographystyle{elsarticle-num}
\bibliography{bib_rohe}

\begin{thebibliography}{10}
\bibitem{ref:alice}
A.~Bean, et~al., The {CMS} pixel detector and challenges (prospectives) for its
  upgrade, these proceedings.
\end{thebibliography}

\end{document}